# On the electric charge of quantized vortices and the dipole moment of vortex pairs and rings in a magnetic field

A. M. Konstantinov and S. I. Shevchenko

*B. Verkin Institute for Low Temperature Physics and Engineering of the National Academy of Sciences of Ukraine*
*Kharkiv 61103, Ukraine*
E-mail: akonstantinov@ilt.kharkov.ua

## Abstract

It is shown that, in the presence of a magnetic field, a quantized vortex line in a superfluid liquid acquires a linear polarization charge, which is localized near the vortex axis over a length on the order of the coherence length. It is found that the total charge of a rectilinear vortex is nonzero, while the vortex pair and vortex ring have a nonzero dipole moment. The electric fields of rectilinear vortices near the end surface of a cylindrical vessel filled with a superfluid liquid are calculated. The electric polarization of superfluid systems in the presence of thermally activated vortex pairs and vortex rings has been studied. It is shown that such a polarization arises in the presence of relative motion of the normal and superfluid components.

## 1. Introduction

An essential property of superfluid and superconducting systems is the possibility of the existence of quantized vortices in them (Onsager-Feynman vortices in superfluid systems and Abrikosov vortices in superconductors). The cardinal difference between Onsager-Feynman vortices and Abrikosov vortices is related to the fact that the Abrikosov vortex carries a magnetic flux $\Phi_0 = \pi \hbar c / e$, where $c$ is the speed of light in vacuum and $e$ is the elementary electric charge. As a result, Abrikosov vortices can be induced by an external magnetic field. They can be observed, for example, by depositing on the superconductor surface small ferromagnetic particles being accumulated at the ends of vortex lines (see, for example, [1]). As shown in [2, 3], a magnetic flux is also associated with Onsager-Feynman vortices, but its value is almost ten orders of magnitude smaller than the flux $\Phi_0$. Therefore, the occurrence of this magnetic flux can be neglected. It is interesting to note that in the presence of a magnetic field, the Onsager-Feynman vortex acquires an electric charge [4, 5]. The compensating charge of the opposite sign arises on the surface of the superfluid system. The charge of the vortex and the compensating charge create an electric field in the surrounding space, whose magnitude can be measured by the currently existing experimental methods. By measuring these fields, one can follow the position and dynamics of vortices and vortex structures, in particular, vortex rings.

Present work is devoted to finding the electric fields generated by quantized vortices in various situations. The article is organized as follows. In Sec. 2 we study general questions related to the polarization of a superfluid system in a magnetic field in the presence of quantized vortex lines, in particular, rectilinear vortices. In Sec. 3 the dipole moment of a vortex pair is found and the polarization of a thin superfluid film in the presence of thermally activated vortex pairs is considered. In Sec. 4, the results of the previous section are generalized to the three-dimensional case: the dipole moment of the vortex ring is calculated and the contribution of thermally activated vortex rings to the polarization of bulk helium is found.

## 2. Vortex filament charge

It is well known that the electric displacement field $\mathbf{D}$ in a dielectric moving in a magnetic field $\mathbf{H}$ with a velocity $\mathbf{v}$ is equal to (see [6] for example)

$$\mathbf{D} = \mathbf{E} + 4\pi \mathbf{P} = \varepsilon \mathbf{E} + \frac{\varepsilon \mu - 1}{c} \mathbf{v} \times \mathbf{H}, \tag{1}$$

where $\varepsilon$ and $\mu$ are permittivity and permeability of the medium respectively. For $^4$He we can assume with good accuracy that $\mu = 1$ and $\varepsilon = 1 + 4\pi n \alpha$ (where $n$ is the system particle density, $\alpha$ is polarizability of helium atoms). In the absence of an external electric field $\mathbf{E}$ from (1) it follows that the polarization $\mathbf{P}$ occurs in the medium:

$$\mathbf{P} = n\alpha \frac{\mathbf{v} \times \mathbf{H}}{c}. \tag{2}$$

Taking into account that the mass flux density is equal to $\mathbf{j} = Mn\mathbf{v}$ ($M$ is the mass of the helium atom), the expression for the polarization can be written in the form

$$\mathbf{P} = \frac{\alpha}{Mc}[\mathbf{j} \times \mathbf{H}]. \tag{3}$$

For a uniform field $\mathbf{H}$ the volume charge density arising in the system is equal to

$$\rho_{pol} = -\operatorname{div} \mathbf{P} = -\frac{\alpha \mathbf{H}}{Mc} \operatorname{rot} \mathbf{j}. \tag{4}$$

In a superfluid liquid in the general case $\mathbf{j} = \rho_n \mathbf{v}_n + \rho_s \mathbf{v}_s$, where $\mathbf{v}_n$ and $\mathbf{v}_s$ are the velocities of the normal and superfluid components, $\rho_n$ and $\rho_s$ are their mass densities. When a quantized vortex line appears in the system, only the superfluid component is moving, so $\rho_n \mathbf{v}_n = 0$. As a result, the flux density is equal to $\mathbf{j} = \rho_s \mathbf{v}_s$. If each point of the vortex line is described by the radius vector $\mathbf{r}_v$, then under assumption that $\rho_s = const$ (taking into account that in this case $\operatorname{rot} \rho_s \mathbf{v}_s = \rho_s \mathbf{\kappa}(\mathbf{r}_v) \delta_2(\mathbf{r} - \mathbf{r}_v)$), we have from (4)

$$\rho_{pol} = -\frac{\alpha \rho_s}{Mc} (\mathbf{\kappa}(\mathbf{r}_v) \cdot \mathbf{H}) \delta_2(\mathbf{r} - \mathbf{r}_v). \tag{5}$$

Here $\delta_2(\mathbf{r} - \mathbf{r}_v)$ is two-dimensional delta function, $\mathbf{\kappa}$ is vortex circulation, the modulus of which is equal to $\kappa = 2\pi\hbar/M$. The direction of vector $\mathbf{\kappa}$ is determined by the tangent to the vortex line at the point $\mathbf{r}_v$, and the motion of the superfluid component seen from the end of the vector is directed counterclockwise. Thus, a linear charge density arises on the vortex axis and it is equal to

$$\lambda(\mathbf{r}_v) = -\frac{\alpha \rho_s}{Mc} (\mathbf{\kappa}(\mathbf{r}_v) \cdot \mathbf{H}). \tag{6}$$

The compensating charge arises on the surface of the system and surface density of this charge is determined by the normal to the surface component of the vector $\mathbf{P}$.

For a rectilinear vortex in a uniform magnetic field, which is directed along the axis of the vortex, the expression (6) takes the form [5]

$$\lambda = -\frac{\alpha \rho_s \kappa s H}{cM}, \tag{7}$$

where $s = \pm 1$ is the sign that determines the direction of circulation of the vortex.

Note that the appearance of charge of vortex in the presence of a magnetic field was also studied for superfluid electron-hole systems [7]. In the cited work, an expression similar to (7) was obtained, in which, however, the polarizability depended on the applied magnetic field. As a consequence, it was found that the magnitude of the emerging electric charge depended on the relationship between the magnetic length and the Bohr radius of electrons and holes. The expression for the linear charge density (6) used in the present work is valid for both two-dimensional (thin films) and bulk superfluid systems, which makes it possible to study the effect not only for rectilinear vortices, but also for vortices of arbitrary shape, in particular, for vortex rings. We also note that in [7] the electric fields created by vortices were not calculated.

Previously, we assumed that $\rho_s = const$. In fact, the density $\rho_s$ depends on the distance to the axis of the vortex and is determined by the dependence on this coordinate of the superfluid velocity $\mathbf{v}_s$. Below in this section, we consider a rectilinear vortex line. In this case, the following velocity field is associated with the vortex:

$$\mathbf{v}_s = \frac{s\hbar}{M} \frac{[\hat{\mathbf{z}} \times \mathbf{r}]}{r^2}. \tag{8}$$

Here, a cylindrical coordinate system is introduced: the axis $z$ of this system coincides with the vortex axis, and $\mathbf{r}$ is the radial component of the radius vector, which describes the distance to the vortex line. Now instead of (5) we can write

$$\rho_{pol} = -\frac{\alpha}{cM} \left( [\mathbf{v}_s \times \mathbf{H}] \nabla \rho_s(r) + \rho_s(r) \mathbf{H} \cdot [\nabla \times \mathbf{v}_s] \right). \tag{9}$$

The vortex related polarization charge (per unit length) $q_{pol}$ can be obtained by integrating (9) over $r$. Taking into account that in the chosen coordinate system $[\nabla \times \mathbf{v}_s]_z = \kappa s \delta_2(\mathbf{r}) = \kappa s \delta(r)/\pi r$ and, as we will show below, $\rho_s(0) = 0$ we find that the integral of the second term in parentheses (9) is equal to zero:

$$\int_0^r \rho_s(r) \mathbf{H} \cdot [\nabla \times \mathbf{v}_s] r dr = \kappa s H \int_0^r \rho_s(r) \frac{\delta(r)}{\pi r} r dr = 0. \tag{10}$$

Thus from (9) we get

$$q_{pol}(r) = 2\pi \int_0^r \rho_{pol} r dr = -\frac{\alpha \kappa s H}{cM} \int_0^r \nabla_r \rho_s(r) dr = -\frac{\alpha \kappa s H}{cM} \rho_s(r). \tag{11}$$

At $r \to \infty$ this charge coincides with (7). An explicit dependence of $\rho_s$ on $r$ can be found in the case of a weakly nonideal bose gas, when the superfluid system is described by the order parameter $\Psi$ satisfying the Gross-Pitaevskii equation. Writing the density of the gas in the form $\rho_s(r) = |\Psi(r)|^2 = \rho_s(\infty) f^2(r)$ (where $f(r)$ is a dimensionless function) from the Gross-Pitaevskii equation we obtain the equation (see [8] for example)

$$\frac{1}{\zeta} \frac{d}{d\zeta} \left( \zeta \frac{df}{d\zeta} \right) - \frac{f}{\zeta^2} + f - f^3 = 0, \tag{12}$$

where $\zeta = r/\xi$, and $\xi$ is the coherence length, which is determined by the pair interaction of atoms. If we introduce the interaction energy of two atoms $U^{(2)}(|\mathbf{r} - \mathbf{r}'|)$, then

$$\xi = \frac{\hbar}{\sqrt{2MU_0 n}}, \quad U_0 = \int U^{(2)}(r) d^3 r. \tag{13}$$

After the numerical solution of the equation (12) and substituting the result into (11), we realize (see Fig. 1), that the charge $q_{pol}$ is localized at several coherence lengths. Below we will assume that the polarization charge density $\rho_{pol}$ is determined by the expression (5), and the superfluid density $\rho_s = const$.

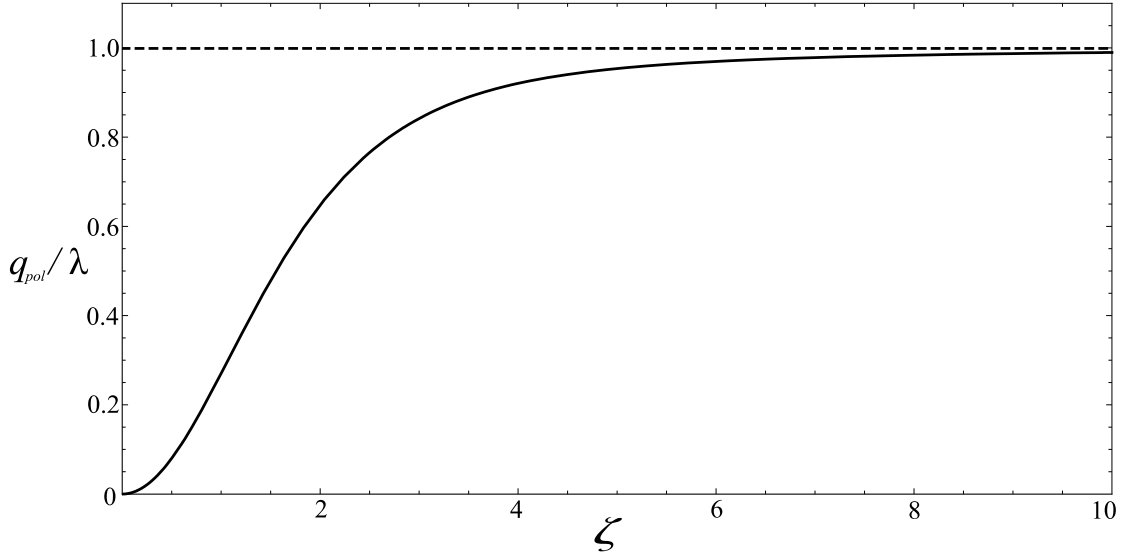

Fig. 1 Charge distribution in a rectilinear vortex.

The emergence of an electric charge on a vortex leads to the appearance of an electric field. The measurement of this field allows to monitor the position of the vortices in the system. It turns out that the appearance of the electric field created by the rectilinear vortices takes place only near the end of the vessel filled with helium [5]. In the case of an infinitely long vessel, the electric field is nonzero only inside the vessel. Indeed, taking into account that in the stationary case $\operatorname{div} \mathbf{v}_s = 0$ and $\operatorname{rot} \mathbf{E} = 0$ from (1) it follows that $\operatorname{rot} \mathbf{D} = (4\pi\alpha\rho_s / Mc)(\mathbf{H} \cdot \nabla) \mathbf{v}_s$. If the magnetic field is directed along the axis of an vortex, which coincides with the axis of the vessel $z$, we discover that if the velocity $\mathbf{v}_s$ is independent on $z$, the vorticity of the vector $\mathbf{D}$ is equal to zero. In the absence of free charges, the divergence of the vector $\mathbf{D}$ is also equal to zero. Accordingly, the electric displacement field will be equal to a constant, which is equal to zero due to the fact that at infinity outside the cylinder $\mathbf{E} = \mathbf{P} = 0$. Thus, if the superfluid velocity $\mathbf{v}_s$ does not depend on the longitudinal coordinate $z$, there is no electric field outside the infinite vessel, and inside the vessel $\mathbf{E} = -4\pi\mathbf{P}$. This field can be expressed in terms of the stream function $\psi$, which determines the velocity field of vortex filaments

$$\mathbf{v}_s = [\hat{\mathbf{z}} \times \nabla]\psi . \qquad (14)$$

In this case, the equation of continuity $\operatorname{div} \mathbf{v}_s = 0$ is fulfilled identically, and the function $\psi$ satisfies the equation

$$\Delta\psi = \sum_i \kappa_i \delta(\mathbf{r} - \mathbf{r}_i), \qquad (15)$$

where $\mathbf{r}_i$ and $\kappa_i$ are the radial coordinate and the circulation of the $i$-th vortex, respectively. Using (3) and (14), we obtain the following expression for the electric field $\mathbf{E} = -(4\pi\alpha\rho_s H / Mc)\nabla\psi$. Taking into account that the potential of the electric field $\varphi$ is connected to the vector $\mathbf{E}$ by the relation $\mathbf{E} = -\nabla\varphi$ we get

$$\varphi = \frac{4\pi\alpha\rho_s H}{Mc}\psi + C . \qquad (16)$$

The constant of integration $C$ is chosen in such a way that the potential on the side surface of the vessel is identically equal to zero (remember that we are now considering an infinitely long vessel).

In the case of a vessel of finite size, the electric potential on the end surface (which we denote as $\varphi_{end}$) is nonzero. For a semi-infinite vessel, the length of which is much greater than its transverse size, the potential $\varphi_{end}$ is two times less than the potential (16). This is easy to see from the integral representation for the potentials. If the end of the semi-infinite vessel lies in the plane $z = 0$, and the transverse dimension of the vessel is much less than the longitudinal one, then

$$\varphi_{end} = \int_{-\infty}^{0} dz' \int d^2 r' \frac{\mathbf{P}(\mathbf{r}') \cdot (\mathbf{r} - \mathbf{r}')}{\left( (\mathbf{r} - \mathbf{r}')^2 + z'^2 \right)^{3/2}}. \tag{17}$$

In the case of an infinite vessel

$$\varphi = \int_{-\infty}^{+\infty} dz' \int d^2 r' \frac{\mathbf{P}(\mathbf{r}') \cdot (\mathbf{r} - \mathbf{r}')}{\left( (\mathbf{r} - \mathbf{r}')^2 + z'^2 \right)^{3/2}}, \tag{18}$$

whence it follows that $\varphi_{end} = \varphi / 2$. The corrections associated with the finite length of the vessel $L$ enter as even degrees of the ratio of the vessel transverse size to the length $L$. For a cylindrical vessel with a circular cross-section of radius $R \ll L$, one can obtain that

$$\varphi_{end} = \int_{-L}^{0} dz' \int d^2 r' \frac{\mathbf{P}(\mathbf{r}') \cdot (\mathbf{r} - \mathbf{r}')}{\left( (\mathbf{r} - \mathbf{r}')^2 + z'^2 \right)^{3/2}} =$$
$$= \varphi / 2 + \frac{\lambda \left( R^2 - r_v^2 \right)}{4L^2} \left\{ -1 + \frac{3}{8L^2} \left( 4r^2 + R^2 + r_v^2 - 4rr_v \cos \theta \right) \right\} \tag{19}$$

Here, a polar coordinate system $(r, \theta)$ is introduced. The origin of this system coincides with the center of the end surface of the cylinder, and the vortex coordinates are equal to $(r_v, 0)$. If these corrections are neglected (the second term in (19)), then the potential $\varphi_{end}$, according to (16), reduces to finding the current function $\psi$. For reference, we write explicit expressions for some current functions.

For a cylindrical vessel with a circular section of the radius $R$, we have (see, for example, [9])

$$\psi = \frac{\kappa}{2\pi} \sum_i \ln \frac{|\mathbf{r} - \mathbf{r}_i|}{|\mathbf{r} - \mathbf{r}_i'|}, \tag{20}$$

where $\mathbf{r}_i' = (R / r_i)^2 \mathbf{r}_i$ is the radial coordinate of the image of $i$-th vortex.

In the case of an annulus ($R_1 < r < R_2$) the stream function is [10]

$$\psi = \frac{\Gamma_1}{2\pi} \ln(r / R_2) + \frac{\kappa}{2\pi} \ln(r / R_2) \sum_j \frac{\ln(R_2 / r_j)}{\ln(R_2 / R_1)} +$$
$$+ \frac{\kappa}{2\pi} \sum_j \operatorname{Re} \ln \left\{ \frac{\vartheta_1 \left( \gamma \ln(r / r_j) + i\gamma (\theta - \theta_j) \right)}{\vartheta_1 \left( \gamma \ln(r_v r / R_2^2) + i\gamma (\theta - \theta_j) \right)} \right\}, \tag{21}$$

where $\Gamma_1$ is the circulation of the velocity vector around the inner cylinder, not associated with vortices, $\gamma \equiv (\pi / 2) \left[ \ln(R_2 / R_1) \right]^{-1}$, $\vartheta_1(x) \equiv \vartheta_1(x \mid 2i\gamma)$ is theta function.

Using the potential $\varphi_{end}$ one can find the transverse components of the electric field on the end of the vessel. However, in order to find the longitudinal component of the field $E_z$, one should use the integral representation of the potential. To do this, we take into account the dependence of the potential (17) on the longitudinal coordinate $z$. Differentiating the expression (17) with respect to $z$, we find

$$E_z = -\partial_z \int_{-\infty}^{0} dz' \int d^2r' \frac{\mathbf{P}(\mathbf{r}') \cdot (\mathbf{r}-\mathbf{r}')}{\left((\mathbf{r}-\mathbf{r}')^2 + (z'-z)^2\right)^{3/2}} = \int d^2r' \frac{\mathbf{P}(\mathbf{r}') \cdot (\mathbf{r}-\mathbf{r}')}{\left((\mathbf{r}-\mathbf{r}')^2 + z^2\right)^{3/2}}. \qquad (22)$$

Let us calculate this field using the example of a cylinder with a circular cross section of radius $R$ and assume that a rectilinear vortex has arisen in the system, the radial coordinate of which is equal to $\mathbf{r}_v = (r_v, 0)$. Integrating (22) by parts and taking into account that $\mathrm{div}\,\mathbf{P} = -\rho_{pol}$ and $P_r(r=R) = \sigma_{pol}$, where $\sigma_{pol}$ is the surface charge density, we find

$$E_z = \frac{\lambda}{\sqrt{r^2 + r_v^2 - 2rr_v \cos\theta + z^2}} + \int_{-\pi}^{\pi} \frac{\sigma_{pol} R d\theta'}{\sqrt{R^2 + r^2 - 2Rr\cos(\theta-\theta') + z^2}} =$$
$$= \lambda \left( \frac{1}{\sqrt{r^2 + r_v^2 - 2rr_v \cos\theta + z^2}} - \frac{1}{\kappa} \int_{-\pi}^{\pi} \frac{v_\theta(R,\theta') R d\theta'}{\sqrt{R^2 + r^2 - 2Rr\cos(\theta-\theta') + z^2}} \right), \qquad (23)$$

where $v_\theta(R,\theta)$ is tangential component of the superfluid velocity on the lateral surface of the cylinder. Taking into consideration that the velocity field $\mathbf{v}_s$ is the sum of the velocity fields of the vortex filament and its "image" located at the point $\mathbf{r}_v = (R^2/r_v, 0)$ we find that the velocity $v_\theta(R,\theta)$ is equal to

$$v_\theta(R,\theta) = \frac{\kappa}{2\pi R} \frac{R^2 - r_v^2}{R^2 + r_v^2 - 2Rr_v \cos\theta}. \qquad (24)$$

Integration in (23) cannot be performed for arbitrary values of $r_v$, so let's consider some special cases.

If the axis of the vortex coincides with the axis of the cylinder ($r_v = 0$), then after substituting (24) into (23) we obtain

$$E_z(r_v = 0) = \lambda \left( \frac{1}{\sqrt{r^2 + z^2}} - \frac{2}{\pi} \frac{K(q)}{\sqrt{(r+R)^2 + z^2}} \right). \qquad (25)$$

Here $K(q)$ is the complete elliptic integral of the first kind and $q^2 = 4rR/((r+R)^2 + z^2)$. The first term in expression (25) is the electric field of a semi-infinite uniformly charged filament, and the second term is the field of a semi-infinite uniformly charged cylinder. At the large distances compared to the radius of the cylinder, the total field decreases as $z^{-3}$. On the end of the vessel (that is, at $z = 0$) near the filament, the field diverges according to a power law (as $1/r$), and near the lateral surface it diverges according to the logarithmic law

$$E_z\left(r_v = 0; |r-R| \ll R; z = 0\right) = \frac{\lambda}{R} - \frac{\lambda}{\pi R} \ln\left(\frac{8R}{|R-r|}\right). \qquad (26)$$

Note that the last statement is valid for an arbitrary location of the vortex.

If the deviation of the vortex from the center of the cylinder is small compared to the radius of the cylinder, then, using the expansion of the velocity (24) of a small parameter $r_v/R$, instead of (25) we have

$$E_z = \lambda \left( \frac{1}{\sqrt{r^2+z^2}} - \frac{2}{\pi} \frac{K(q)}{\sqrt{(r+R)^2+z^2}} \right) +$$
$$+ \frac{\lambda r_v \cos\theta}{R} \left( \frac{rR}{(r^2+z^2)^{3/2}} - \frac{2}{\pi} \frac{(r^2+R^2+z^2)K(q)-((r+R)^2+z^2)E(q)}{rR\sqrt{(r+R)^2+z^2}} \right). \quad (27)$$

Here $E(q)$ is the complete elliptic integral of the second kind.

If the distance from the vortex to the wall $R-r_v$ is small compared to the radius of the cylinder $R$, then the compensating surface charge is concentrated in a small region near the cylinder surface, the polar angles of which are limited by the interval $\theta \in (-(R-r_v)/R; (R-r_v)/R)$. Therefore, it can be asserted that the electric field in this case is equivalent to the field of two oppositely charged semi-infinite filaments located at the distance of the order $R-r_v$ from each other. In this case, the product $\lambda(R-r_v)$ will play a role of the linear density of the dipole moment with an arm equal to the distance from the vortex line to the lateral surface of the cylindrical vessel.

In conclusion, we give an estimate of the magnitude of the electric field $E_z$. For liquid helium (whose density per unit mass is $\rho_s/M = 2\cdot 10^{22} cm^{-3}$, and polarizability $\alpha = 2\cdot 10^{-25} cm^3$) in a magnetic field $10^5 G$ the value of $E_z$ is of the order of $4\cdot 10^{-9} V/cm$.

As is known, rectilinear vortices are generated in a cylinder when it is rotated with an angular velocity $\Omega$ exceeding the critical value $\Omega_c = (\hbar/MR^2)\ln(R/\xi)$. When the rate of rotation of the cylindrical vessel increases further, new vortex filaments appear, and when $\Omega \gg \Omega_c$ their number is very large. In this limit they simulate the rotation of the superfluid part of the liquid as a rigid body with a velocity $\mathbf{v}_s = \mathbf{\Omega}\times\mathbf{r}$. Using this velocity in (2), with the help of (22) one can obtain an expression for the resulting field $E_z$ created by the vortex lattice:

$$E_z = \frac{4\rho_s \alpha H \Omega}{Mc} \left( \frac{R^2 K(q)}{\sqrt{(r+R)^2+z^2}} - 2\int_0^R \frac{r' K(q')}{\sqrt{(r+r')^2+z^2}} dr' \right), \quad (28)$$

where $q'^2 = 4rr'/((r+r')^2+z^2)$. At the large distances compared to the radius of the cylinder, the total field decreases according to the cubic law:

$$E_z(r \sim z \gg R) = \frac{\pi \rho_s \alpha H \Omega}{4Mc} \frac{R^4(r^2-2z^2)}{(r^2+z^2)^{5/2}}. \quad (29)$$

On the axis of the cylinder (at $r = 0$), the field (28) reduces to the expression

$$E_z(r=0) = \frac{2\pi \rho_s \alpha H \Omega}{Mc} \left( 2z - \frac{R^2+2z^2}{\sqrt{R^2+z^2}} \right). \quad (30)$$

For $z = 0$, this expression determines the maximum value of the field (28). For helium filling a cylinder of radius $R = 1 cm$ at the angular velocity of its rotation $\Omega = 10 s^{-1}$ and magnetic field $10^5 G$, the magnitude of this field $E_z$ can reach the order of $2.5\cdot 10^{-4} V/cm$.

3. **Dipole moment of vortex pairs**

Let us now consider superfluid helium film in which a vortex pair has arisen, that is, two vortices of opposite circulation spaced apart from each other at a distance **l**. Taking into account that the dipole moment is a vector, at the ends of which there are charges of the opposite sign, we find from (7) the following expression for the dipole moment of the vortex pair

$$\mathbf{d}_v = -\frac{\alpha \rho_s H}{cM}\frac{2\pi\hbar}{M}\mathbf{l}, \qquad (31)$$

In this section $\rho_s$ is the two-dimensional density of the superfluid component, proportional to the film thickness $h$, and $\mathbf{l}$ is a vector connecting the negative circulation vortex with the positive circulation vortex, and it is assumed that the vector $\mathbf{l}$ is parallel to the substrate. It is also possible to verify the validity of (31) by directly integrating the polarization (3). To do this, we will write again the mass flux density in the form $\mathbf{j} = \rho_s \mathbf{v}_{sv}$, where $\mathbf{v}_{sv}$ is the velocity field created by the vortex pair at the point $\mathbf{r}$. In order to find the dipole moment of the system, one should integrate the velocity $\mathbf{v}_{sv}$ over the area of the system $S$. For this, it is convenient to write the velocity $\mathbf{v}_{sv}$ in the form $\mathbf{v}_{sr} = \hbar\nabla\phi/M$ (where $\phi$ is the phase of the order parameter). As a result, we arrive at the contour integral

$$\int \mathbf{v}_{sv}(\mathbf{r})dS = \frac{\hbar}{M}\oint \phi\, d\Sigma, \qquad (32)$$

where $d\Sigma$ is the normal to the integration contour, numerically equal to the length element of the contour. It should be borne in mind that the phase is an ambiguous function of the coordinates and therefore it is necessary to make a cut between the vortices of the pair, as shown in Fig. 2

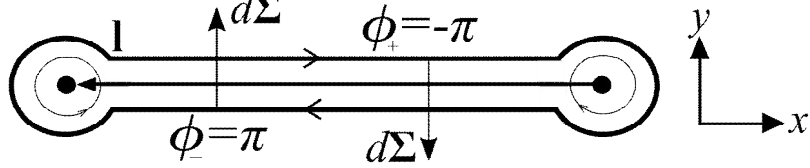

Fig. 2 The integration contour used in the calculation of the dipole moment of the vortex pair.

On opposite sides of the cut, the phase differs by $2\pi$. Let us denote the values of the phase $\phi$ on the upper and lower sides of the cut by $\phi_+$ and $\phi_-$, respectively, and taking into account that the contributions to the integral (32) from the cores of the vortex pair compensate each other, for an unbounded system we obtain

$$\int \mathbf{v}_{sv}(\mathbf{r})dS = \frac{\hbar}{M}\int_0^l (-\phi_+ + \phi_-)\hat{\mathbf{y}}dx = -\frac{2\pi\hbar}{M}\hat{\mathbf{z}}\times\mathbf{l}. \qquad (33)$$

Here $\hat{\mathbf{z}}$ is the unit vector directed along the normal to the substrate. It follows from (3) and (33) that the dipole moment is equal to

$$\mathbf{d}_v = \int \mathbf{P}dS = -\frac{\alpha\rho_s H}{cM}\frac{2\pi\hbar}{M}\left[[\hat{\mathbf{z}}\times\mathbf{l}]\times\mathbf{H}\right] = -\frac{\alpha\rho_s H}{cM}\frac{2\pi\hbar}{M}\mathbf{l}, \qquad (34)$$

which coincides with (31).

This expression can be rewritten in terms of the momentum of the vortex pair $\mathbf{p}$. In order to do this, we write down the energy of a vortex pair, which consists of the kinetic energy of the superfluid component in the entire space (except for the cores of vortices) and the energy of the cores of vortices. Taking into account that outside the cores the density of the superfluid component is a constant and denoting the energy of the vortex core by $\Delta$, we find the total energy of the vortex pair

$$E_{0v} = \frac{\rho_s}{2}\int v_{sv}^2 dS + 2\Delta. \qquad (35)$$

Expressing again the velocity $\mathbf{v}_{sv}$ in terms of the phase $\phi$, we rewrite (31) in the form

$$E_{0v} = \frac{\hbar^2 \rho_s}{2M^2} \int \left[ \nabla(\phi \nabla \phi) - \phi \nabla^2 \phi \right] dS + 2\Delta. \qquad (36)$$

The second term in square brackets contains $\nabla^2 \phi = (M/\hbar) \text{div } \mathbf{v}_{sv}$ and therefore in the stationary case it is equal to zero. The first term is reduced to the contour integral and the expression (36) takes the form

$$E_{0v} = \frac{\hbar^2 \rho_s}{2M^2} \oint \phi \nabla \phi d\Sigma + 2\Delta. \qquad (37)$$

It can be seen that the only contribution to (37) arises due to the cut (see Fig. 2). When calculating the integral (37) along the cut, it should be taken into account that the product $\phi d\Sigma$ is the same on both sides of the cut. Let us choose the center of the positive circulation vortex as the origin of coordinates and direct the axis $x$ along the line connecting the vortices. As a result, the energy $E_{0v}$ takes the form

$$E_{0v} = \frac{\pi \hbar \rho_s}{M} \int_{\xi}^{l-\xi} \mathbf{v}_{sv} \cdot \hat{\mathbf{y}} dx + 2\Delta, \qquad (38)$$

where $\xi$ is now the radius of the vortex core, which size is of the order of the coherence length. The velocity field of the vortex pair in the chosen coordinate system has the form (compare with (14) and (20))

$$\mathbf{v}_{sv}(\mathbf{r}) = \frac{\hbar}{M} \hat{\mathbf{z}} \times \nabla \ln \frac{r}{|\mathbf{r} - \mathbf{r}_-|}. \qquad (39)$$

Here $\mathbf{r}_-$ is the radius vector of the negative circulation vortex. Using (38) and (39), we find that

$$E_{0v} = \frac{\pi \hbar^2 \rho_s}{M^2} \int_{\xi}^{l-\xi} [\hat{\mathbf{y}} \times \hat{\mathbf{z}}] \cdot \nabla \ln \left| \frac{x}{x - x_-} \right| dx + 2\Delta. \qquad (40)$$

Assuming that $l \gg \xi$, the final expression for the energy of the vortex pair takes the form

$$E_{0v} = \frac{2\pi \rho_s \hbar^2}{M^2} \ln \frac{l}{\xi} + 2\Delta. \qquad (41)$$

According to Kelvin's theorem, quantized vortices move at a speed equal to the local superfluid velocity at the locations of the vortices. For a vortex pair, this velocity is the same for both vortices and is equal to

$$\mathbf{v}_L = -\frac{\hbar}{M} \frac{\hat{\mathbf{z}} \times \mathbf{l}}{l^2}. \qquad (42)$$

From this expression it follows that the vortex pair moves as a whole with a velocity $\mathbf{v}_L$ in the direction normal to the line connecting the vortex centers. Knowing the energy of the pair and its velocity, we can introduce the momentum of the pair using the relation $\mathbf{v}_L = dE_{0v}/d\mathbf{p}$. Integrating this expression and setting the integration constant equal to zero (because there is no vortex pair at $l = 0$), we find that the momentum of the vortex pair is equal to

$$\mathbf{p} = -\frac{2\pi \hbar \rho_s}{M} [\hat{\mathbf{z}} \times \mathbf{l}]. \qquad (43)$$

This momentum is equal to the total flow of the superfluid component created by the pair in the entire system: $\mathbf{p} = \rho_s \int \mathbf{v}_{sv} dS$ (see (33)). From (34) and (43) we get

$$\mathbf{d}_v = \frac{\alpha}{Mc} [\mathbf{p} \times \mathbf{H}]. \qquad (44)$$

In the absence of an average current, the pairs are randomly oriented in space and their average dipole moment is equal to zero. In the presence of a current, the situation changes and the value of the vector $\mathbf{d}_v$ (per unit area) averaged over the positions and orientations of the pairs is equal to

$$\langle \mathbf{P}_v \rangle \equiv \frac{1}{S}\int \mathbf{d}_v \Gamma(\mathbf{l},t)d^2r d^2l = -\frac{\alpha \rho_s H}{cM}\frac{2\pi\hbar}{M}\int \mathbf{l}\Gamma(\mathbf{l},t)d^2l. \tag{45}$$

Here the function $\Gamma(\mathbf{l},t)$ is the number of pairs with the size $\mathbf{l}$ per unit area in the volume element $d^2l$. If the substrate (which the superfluid film is coating) is rough and the movement of vortex pairs along the substrate occurs randomly, the function $\Gamma(\mathbf{l},t)$ satisfies the Fokker-Planck equation (see, for example, [11])

$$\frac{\partial \Gamma}{\partial t} = 2D\frac{\partial}{\partial \mathbf{l}}\left(\frac{1}{T}\frac{\partial E_v}{\partial \mathbf{l}}\Gamma + \frac{\partial \Gamma}{\partial \mathbf{l}}\right), \tag{46}$$

where $D$ is the diffusion coefficient, $E_v$ is the energy of vortex pairs minus the product of momentum $\mathbf{p}$ and relative velocity $\mathbf{w} = \mathbf{v}_n - \mathbf{v}_{s0}$ (where $\mathbf{v}_n$ and $\mathbf{v}_{s0}$ are the velocities of the normal and superfluid components of the external flow). This energy is equal to [12]

$$E_v = \frac{2\pi \rho_s \hbar^2}{M^2}\int_\xi^l \frac{dr}{r\tilde{\varepsilon}(r)} + 2\Delta - \mathbf{p}\cdot\mathbf{w}. \tag{47}$$

This expression contains a scale-dependent "dielectric constant" $\tilde{\varepsilon}(r)$ introduced by Kosterlitz and Thouless. It effectively takes into account the attenuating effect of smaller pairs on the interaction between the respective members of larger pairs (compare (41) and (47)).

On the other hand, the average flux of a superfluid liquid associated with vortex pairs is equal to

$$\langle \mathbf{j}_v \rangle \equiv \int \rho_s \mathbf{v}_{sv}(\mathbf{r})\Gamma(\mathbf{l},t)d^2l d^2r = -\frac{2\pi\hbar\rho_s}{M}\hat{\mathbf{z}}\times \int \mathbf{l}\Gamma(\mathbf{l},t)d^2l, \tag{48}$$

where $\mathbf{v}_{sv}(\mathbf{r})$ velocity field of the superfluid component created by the vortex pair. Note that integration over $r$ in (48) uses (33). From (45) and (48) it follows that

$$\langle \mathbf{P}_v \rangle = -\frac{\alpha H}{cM}\left[\hat{\mathbf{z}}\times \langle \mathbf{j}_v \rangle\right]. \tag{49}$$

The presence of vortices renormalizes the superfluid density. The renormalized superfluid density $\mathbf{p}_r = (2\pi\hbar/M)\rho_s \pi r_c^2 \mathbf{e}_p$ can be introduced using the relation for the total mass flux $\mathbf{j}$

$$\mathbf{j} = \rho_s \mathbf{v}_s + \langle \mathbf{j}_v \rangle + (\rho - \rho_s)\mathbf{v}_n \equiv \rho_{sv}\mathbf{v}_{s0} + (\rho - \rho_{sv})\mathbf{v}_n, \tag{50}$$

where $\rho$ is the total mass density of the liquid and $\rho_s$ is the mass superfluid density of the liquid in the absence of vortices. Using the definition of "dielectric constant" $\tilde{\varepsilon}(r\to\infty,T)\equiv \epsilon(T) = \rho_s(T)/\rho_{sv}(T)$ ($T$ is the temperature) from (49) and (50) we finally find

$$\langle \mathbf{P}_v \rangle = -\frac{\alpha \rho_s H}{cM}(1-\epsilon^{-1})[\hat{\mathbf{z}}\times \mathbf{w}]. \tag{51}$$

The expression for polarization (51) at small $\epsilon - 1$ can be obtained by direct integration of (45). Indeed, the stationary solution of Eq. (46) has the form

$$\Gamma = \frac{1}{\xi^4}\exp\left[-\frac{E_v}{T}\right]. \tag{52}$$

Using (47) and (52) in (45) in the linear approximation in $\mathbf{w}$ we obtain

$$\langle \mathbf{P}_v \rangle = -\frac{\alpha \rho_s H}{cM} [\hat{\mathbf{z}} \times \mathbf{w}] \cdot A, \tag{53}$$

where the constant $A$ is defined by the following expression

$$A = \frac{4\pi^3 \hbar^2 \rho_s \exp[-2\Delta/T]}{M^2 T} \int_\xi^\infty \left( \frac{l^3}{\xi^4} \exp\left[ -\frac{2\pi \hbar^2 \rho_s}{M^2 T} \int_\xi^l \frac{dr'}{r' \tilde{\varepsilon}(r')} \right] \right) dl. \tag{54}$$

On the other hand, the known Kosterlitz-Thouless recurrence relation for the "dielectric constant" $\tilde{\varepsilon}(r)$ in the limit $r \to \infty$ takes the form $\epsilon = 1 + A$ (see [12, 13]) so that for the polarization (53) we finally have

$$\langle \mathbf{P}_v \rangle = -\frac{\alpha \rho_s H}{cM} (\epsilon(T) - 1)[\hat{\mathbf{z}} \times \mathbf{w}]. \tag{55}$$

In conclusion, we note that while obtaining the polarization $\langle \mathbf{P}_v \rangle$ it was assumed that the velocity of the relative motion of the normal and superfluid components $\mathbf{w}$ is constant. However, if $\mathbf{w}$ is a slow function of the coordinates, then expression (51) can be used to calculate the electric potential created by the vortex pairs in the surrounding space. According to the standard formula of electrostatics, this potential is equal to

$$\varphi(\mathbf{r}, z) = \int d^2 r' \frac{\langle \mathbf{P}_v \rangle \cdot (\mathbf{r} - \mathbf{r}')}{\left((\mathbf{r} - \mathbf{r}')^2 + z^2\right)^{3/2}} =$$
$$= -\frac{\alpha \rho_s H}{cM} (1 - \epsilon^{-1}) \int d^2 r' \frac{[\mathbf{w}(\mathbf{r}') \times (\mathbf{r} - \mathbf{r}')]_z}{\left((\mathbf{r} - \mathbf{r}')^2 + z^2\right)^{3/2}}, \tag{56}$$

Let the film occupy the wide strip $-W/2 \ll x \ll W/2$ and $-\infty < y < \infty$. Then for a constant velocity of relative motion $\mathbf{w}_0$ directed along the axis $y$ from (56) it follows that the potential is equal to

$$\varphi(\mathbf{r}, z) = \frac{2\alpha \rho_s H w_0}{cM} (1 - \epsilon^{-1}) \frac{xW}{(W/2)^2 + z^2}. \tag{57}$$

If a plane surface wave (the third sound) propagates along the axis in such a way that $w(y, t) = w_a \exp(i\omega t - iky)$ (where $w_a$, $\omega$ and $k$ are the amplitude, frequency, and wave number, respectively), we find from (56) that

$$\varphi(\mathbf{r}, z) = \frac{\alpha \rho_s H w(y, t)}{cM} (1 - \epsilon^{-1}) \frac{xWk}{\sqrt{(W/2)^2 + z^2}} K_0 \left( k \sqrt{(W/2)^2 + z^2} \right), \tag{58}$$

where $K_0(x)$ is the modified Bessel function of the second kind. For a saturated liquid helium film with a thickness $h = 10^{-6} cm$ placed in a magnetic field $10^5 G$, the value of the potential $\varphi$ at a relative speed $w = 1 cm/s$ reaches the order of $4 \cdot 10^{-12} (1 - \epsilon^{-1}) V$. The difference $1 - \epsilon^{-1}$ included in this expression is limited by the interval $1 - \epsilon^{-1} \in [0; 1)$. Recalling that the Kosterlitz-Thouless "permittivity" $\epsilon = \rho_s(T) / \rho_{sv}(T)$ is an increasing function of temperature, it can be argued that the difference $1 - \epsilon^{-1}$ (and, accordingly, the value of the potential $\varphi$) also increases significantly with temperature.

## 4. Dipole moment of vortex rings

Let us consider a bulk superfluid liquid in which a quantized vortex ring of radius $r_c$ has appeared. In this section, we will talk about unbounded liquid. In the presence of a uniform magnetic field $\mathbf{H}$, according to (5), a polarization charge density will arise on the ring:

$$\rho_{pol} = -\frac{\alpha \rho_s}{Mc}\frac{2\pi\hbar}{M}\left(\mathbf{H}\cdot\hat{\boldsymbol{\theta}}\right)\delta(r-r_c)\delta(z). \tag{59}$$

Here now $\rho_s$ is the three-dimensional density of the superfluid component. In (59) a cylindrical coordinate system is introduced, whose origin coincides with the center of the ring, the unit vector $\hat{\mathbf{z}}$ is directed along the rotation axis of the ring, perpendicular to its plane, and the polar unit vector $\hat{\boldsymbol{\theta}} = [\hat{\mathbf{z}}\times\mathbf{r}]/r$ is directed along the circulation vector of the ring $\boldsymbol{\kappa}$. From (59) it follows that the vortex ring has no total charge. However, its dipole moment $\mathbf{d}_r$ is nonzero. Denoting the radius vector of the points of the system by $\mathbf{R}$, we write this dipole moment in the following form:

$$\mathbf{d}_r = \int \rho_{pol}\mathbf{R}dV = \frac{\alpha\rho_s}{Mc}\frac{2\pi\hbar}{M}\int_0^{2\pi}\left(\mathbf{r}\cdot[\hat{\mathbf{z}}\times\mathbf{H}]\right)rd\theta\bigg|_{r=r_c} = \frac{2\pi^2\hbar\alpha\rho_s r_c^2}{M^2 c}[\hat{\mathbf{z}}\times\mathbf{H}]. \tag{60}$$

This expression can also be obtained by direct integration in (3). Indeed, taking into account that in the case under consideration the mass flux $\mathbf{j} = \rho\mathbf{v} = \rho_s\mathbf{v}_{sr}$ (where $\mathbf{v}_{sr} = \hbar\nabla\phi/M$ is the velocity field created by the vortex ring), we obtain

$$\mathbf{d}_r = \int \mathbf{P}dV = -\frac{\alpha\rho_s\hbar}{M^2 c}\mathbf{H}\times\int\phi d\mathbf{S}_r. \tag{61}$$

where $d\mathbf{S}_r$ is the area element to which the direction of the normal to the integration surface is assigned. The phase $\phi$ entering into (61) is an ambiguous function of the coordinates. In order for the phase to be determined uniquely, from the volume over which the integration is performed, one should exclude a cylindrical region, whose height and radius are of the order of the transverse size of the vortex core and the radius of the vortex ring, respectively, and the center coincides with the center of the vortex ring (see Fig. 3)

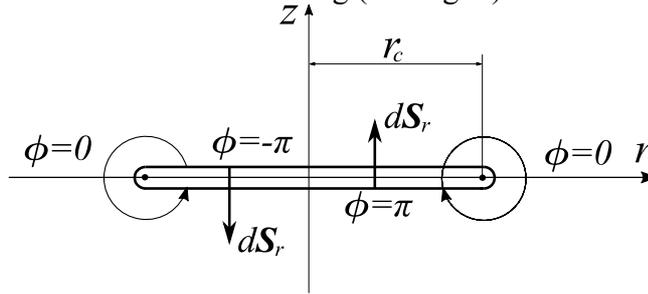

Fig. 3 Cross section of the integration surface used in calculating the dipole moment of the vortex ring.

Using the fact that $\phi(r>r_c, z=0)=0$ and $\phi(r<r_c, z=\pm 0)=\mp\pi$, we find that the integral of the phase $\phi$ over the circular surface $\mathbf{S}_r$ in (61) is equal to $2\pi^2 r_c^2\hat{\mathbf{z}}$. Note that the direction of this quantity coincides with the direction in which the ring would move in an infinite fluid (hereinafter we will also refer to this direction as $\mathbf{e}_p$). As a result, we come again to (60). Taking into account that the momentum of the vortex ring is equal to $\mathbf{p}_r = (2\pi\hbar/M)\rho_s\pi r_c^2\hat{\mathbf{z}}$ (see, for example, [9]), its dipole moment can be rewritten as

$$\mathbf{d}_r = \frac{\alpha}{Mc}[\mathbf{p}_r\times\mathbf{H}]. \tag{62}$$

This expression is analogous to the dipole moment of the vortex pair (44) in the two-dimensional case.

Similar to the case of vortex pairs considered above, the average polarization $\langle \mathbf{P}_r \rangle$ associated with vortex rings will be nonzero in the presence of the external flow. In order to find $\langle \mathbf{P}_r \rangle$, we use expression (45), in which we should proceed to integration over the size and orientation of the ring. In this case, the function $\Gamma(\mathbf{l},t)$ should be replaced by the function $\Gamma_r(\mathbf{r}_c,t)$, which is the number density of vortex rings of radius $r_c$ with orientation $\mathbf{e}_p$ per unit volume $d^3 r_c$ so that

$$\langle \mathbf{P}_r \rangle = \int \mathbf{d}_r \Gamma_r(\mathbf{r}_c,t) d^3 r_c . \tag{63}$$

In the stationary case, the function $\Gamma_r(\mathbf{r}_c,t)$ is determined by the expression $\Gamma_r = (1/\xi^6)\exp\left[-E_r(\mathbf{r}_c)/T\right]$, where $E_r(\mathbf{r}_c)$ is the energy of the vortex ring of radius $r_c$. In order to go further, we use the results of works [14-16], which are devoted to the generalization of the Kosterlitz-Thouless theory to the three-dimensional case. The main ideas of these works are as follows.

The presence of thermally activated vortices in the system should lead to a renormalization of the superfluid density. This follows from the fact that the applied external flow will be partially screened by the current resulting from the oriented vortex rings, which, in turn, will lead to a decrease in the density of the superfluid liquid. In addition, similarly to the two-dimensional case, one should take into account the renormalization of the vortex-rings energy by rings of smaller radius. The unscreened energy of one vortex ring is equal to

$$E_{0r} = \frac{2\pi^2 r_c \rho_s \hbar^2}{M^2}\left(\ln(r_c/\xi) + C\right) + \Delta, \tag{64}$$

where $\Delta$ is the energy needed to form a vortex ring of radius $\xi$ and $C$ is a constant related to the energy $\Delta$, whose value for superfluid helium is $C = 0.464$ [17]. The value $\xi$ here is some suitable cutoff to avoid spurious divergences at small separations, and expression (64) is valid for thin vortex rings, that is, for $r_c \gg \xi$. With logarithmic accuracy, the energy (64) can be obtained from the expression for the energy of the vortex pair (41) with the pair size replaced by the ring radius ($l \to r_c$) and with the two-dimensional density of the superfluid component replaced by the three-dimensional density multiplied by half the ring length ($\rho_s \to \pi r_c \rho_s$). The resulting effective ring energy is obtained by dividing the energy $E_{0r}$ of a vortex ring by a scale-dependent constant $\tilde{\mu}(r_c)$. In the presence of relative velocity $\mathbf{w}$, we have

$$E_r = \frac{2\pi^2 \rho_s \hbar^2}{M^2} \int_\xi^{r_c} \frac{\ln(r/\xi) + C + 1}{\tilde{\mu}(r)} dr - \mathbf{w}\cdot\mathbf{p}_r + \Delta. \tag{65}$$

In order to find a self-consistent equation for the scale-dependent constant $\tilde{\mu}(r_c)$, the magnetostatic analogy was exploited in which vortex rings were regarded as dipole current loops, having a dipole moment $\mathbf{m}(r_c) = (\pi \hbar r_c^2 / 2M)\mathbf{e}_p$ (see also [18]). In this case, $\tilde{\mu}(r_c)$ has the meaning of the magnetic permeability, which is related to the magnetic susceptibility $\chi_m(r_c)$ by the well-known relation $\tilde{\mu}(r_c) = 1 + 4\pi \chi_m(r_c)$. The magnetic susceptibility, in its turn, is determined by the magnetic polarizability $\alpha_m(r_c)$:

$$\chi_m(r_c) = \int_\xi^{r_c} \alpha_m(r') dn(r'). \tag{66}$$

where $n(r)$ is the number density of rings and is related to the $\Gamma_r$ by the relation $dn(r') = \Gamma_r|_{w=0} d^3r'$. From (66) one can find the magnetic permeability $\tilde{\mu}(r_c)$ once $\alpha_m(r_c)$ and $\Gamma_r|_{w=0}$ are known.

The magnetic polarizability of a vortex ring is given by the expression

$$\alpha_m(r_c) = \frac{\partial}{\partial w}\langle m\cos\theta\rangle\bigg|_{w=0}, \tag{67}$$

where $\theta$ is the angle between the magnetic moment $\mathbf{m}$ and the relative velocity $\mathbf{w}$, and the averaging is done over the orientation of the vortex ring. It can be seen from (65) and (67) that the relative velocity plays the role of an external magnetic field $\mathbf{B}$ so we can use the following transcription: $\mathbf{mB} \leftrightarrow (2\pi\hbar/M)\rho_s \pi r_c^2 (\mathbf{w}\cdot\mathbf{e}_p)$. In the case of a circular vortex ring, from (65) and (67) we obtain

$$\alpha_m(r_c) = \frac{\int d\Omega m(r_c)\cos\theta\left(\frac{\partial}{\partial w}\exp(-E_r/T)\right)\bigg|_{w=0}}{\int d\Omega \exp(-E_r/T)\big|_{w=0}} = \frac{2m(r_c)p_r}{3T} = \left(\frac{\pi^3\hbar^2\rho_s}{3M^2 T}\right)r_c^4. \tag{68}$$

Here $d\Omega = r_c^2 \sin\theta d\theta d\varphi$ is the solid angle element. As a result, from (66) and (68) for the scale-dependent constant $\tilde{\mu}(r_c)$ follows the recursion relation

$$\tilde{\mu}(r_c) = 1 + \\ + \frac{16\pi^5\hbar^2\rho_s \exp[-\Delta/T]}{3M^2 T}\int_\xi^{r_c}\left(\frac{r^6}{\xi^6}\exp\left[-\frac{2\pi^2\hbar^2\rho_s}{M^2 T}\int_\xi^r \frac{\ln(r'/\xi)+C+1}{\tilde{\mu}(r')}dr'\right]\right)dr, \tag{69}$$

and in the theory under consideration $\tilde{\mu}(r_c \to \infty) \equiv \mu_r(T) = \rho_s/\rho_{sr}$, where $\rho_{sr}$ is the superfluid density in the presence of thermally activated vortex rings. We emphasize that the quantities appearing here and are introduced by magnetostatic analogy, but they should not be confused with the true magnetic permeability of the medium. It is worth emphasizing that the quantities $\tilde{\mu}$ and $\mu_r$ appearing here are introduced by magnetostatic analogy, but they should not be wrongly associated with the genuine magnetic permeability of the medium $\mu$ (see Sec. 1). At low temperatures, the integral in (69) is small and the value of $\tilde{\mu}(r_c)$ is close to unity, so that in the integrand in (69) the magnetic permeability $\tilde{\mu}(r')$ can be replaced by unity. At high temperatures near the phase transition $T_c$, the quantity $\mu_r(T)$ as a function of relative temperature $T/(T_c - T)$ diverges according to a power law. For circular vortex rings, the exponent is equal to 0.57 (see [16]).

Returning to the averaging of the dipole moment of system (63), we note that the average polarization $\langle \mathbf{P}_r \rangle$ can be related to the average flux of a superfluid liquid $\langle \mathbf{j}_r \rangle$, created by vortex rings. Indeed, substituting into (63) the dipole moment of the ring (62), expressed in terms of momentum, we find

$$\langle \mathbf{P}_r \rangle = -\frac{\alpha \mathbf{H}}{Mc}\times \int \mathbf{p}_r \Gamma_r(\mathbf{r}_c,t)d^3 r_c. \tag{70}$$

On the other hand, the flux $\langle \mathbf{j}_r \rangle$ is determined by the expression

$$\langle \mathbf{j}_r \rangle \equiv \int \rho_s \mathbf{v}_{sr}(\mathbf{r})\Gamma_r(\mathbf{r}_c,t)d^3 r_c d^3 r = \int \mathbf{p}_r \Gamma_r(\mathbf{r}_c,t)d^3 r_c. \tag{71}$$

Thus,

$$\langle \mathbf{P}_r \rangle = -\frac{\alpha}{Mc}\left[\mathbf{H} \times \langle \mathbf{j}_r \rangle\right], \tag{72}$$

which formally coincides with expression (49). By analogy with the two-dimensional case, the superfluid density $\rho_{sr}$ in the presence of vortex rings is introduced using the relation for the total mass flux $\mathbf{j}$

$$\mathbf{j} = \rho_s \mathbf{v}_{s0} + \langle \mathbf{j}_r \rangle + (\rho - \rho_s)\mathbf{v}_n \equiv \rho_{sr}\mathbf{v}_{s0} + (\rho - \rho_{sr})\mathbf{v}_n. \tag{73}$$

Combining (72) and (73), we finally find

$$\langle \mathbf{P}_r \rangle = -\frac{\alpha \rho_s}{cM}\left(1 - \mu_r^{-1}\right)\left[\mathbf{H} \times \mathbf{w}\right]. \tag{74}$$

As in the two-dimensional case, for small $\mu_r - 1$ one can arrive at (74) directly by calculating the expression (63). Expanding the function $\Gamma_r$ in terms of the relative velocity $\mathbf{w}$ and taking into account the relation (69), in the linear approximation in $\mathbf{w}$ we obtain

$$\langle \mathbf{P}_r \rangle = -\frac{\alpha \rho_s}{cM}\left[\mathbf{H} \times \mathbf{w}\right]\cdot\left(\mu_r(T) - 1\right). \tag{75}$$

For liquid helium in a magnetic field $10^5 G$ and at relative velocity $w = 1 cm/s$, the polarization $\langle \mathbf{P}_r \rangle$ is equal to $4\cdot 10^{-6}\left(1 - \mu_r^{-1}\right) V/cm$.

## 5. Conclusions

In this paper, we have demonstrated that in the presence of a magnetic field, a quantized vortex line in a superfluid liquid acquires a linear charge density. As a consequence, it has been found that a polarization charge arises in a rectilinear vortex, while a dipole moment arises in a vortex pair and a vortex ring. The observation of electric fields created by vortex structures makes it possible to track their position in a superfluid liquid. In particular, for rectilinear vortices such an observation can be carried out near the end surface of a cylinder filled with helium, and for vortex pairs, over a substrate covered with a superfluid liquid. In the paper we have also considered a system of thermally activated vortex pairs (for the two-dimensional case) and vortex rings (for the three-dimensional case) and it has been shown that their assembly leads to polarization of the liquid in the presence of relative motion of the normal and superfluid components.

It is worth noting that the relative motion in a superfluid liquid can be caused, for example, by a temperature gradient that induces a heat flux. In the presence of a magnetic field, this flux leads to the appearance of an electric field in the surrounding space [19, 20]. The presence of thermally activated vortex structures in the liquid will lead to additional polarization of the system, which, however, will not qualitatively change the effect predicted in [19, 20]. A noticeable contribution of vortices can be detected if the difference $1 - \mu_r^{-1}$ is not small, that is, near the superfluid transition temperature.

It should also be noted that, in the effect predicted in [19, 20], the magnitude of the emerging electric field depends significantly on the shape of the sample with helium. In particular, for a circular cylinder in the absence of quantized vortices, an electric field does not arise outside the cylinder. Therefore, the fields obtained in the work associated with quantized vortex structures can serve as an indicator of their creation and, for example, the transition of a superfluid liquid to a turbulent state.